\documentclass[10pt]{article}

\usepackage[OE]{express}

\begin{document}
\title{Critical side channel effects in random bit generation with multiple semiconductor lasers in a polarization-based quantum key distribution system}
\author{Heasin Ko,\authormark{1$\dagger$} Byung-Seok Choi,\authormark{1} Joong-Seon Choe,\authormark{1} Kap-Joong Kim,\authormark{1} Jong-Hoi Kim,\authormark{1} and Chun Ju Youn\authormark{1,2*}}
\address{\authormark{1}Photonic/Wireless Convergence Components Research Division, Electronics and Telecommunications Research Institute, Daejeon, 34129, South Korea\\}
\address{\authormark{2}School of Advanced Device Technology, University of Science \& Technology, Daejeon, 34113, South Korea\\}
\email{\authormark{*}cjyoun@etri.re.kr,\authormark{$\dagger$}seagod.ko@etri.re.kr} 


\begin{abstract}
Most polarization-based BB84 quantum key distribution (QKD) systems utilize multiple lasers to generate one of four polarization quantum states randomly. However, random bit generation with multiple lasers can potentially open critical side channels, which significantly endangers the security of QKD systems. In this paper, we show unnoticed side channels of temporal disparity and intensity fluctuation, which possibly exist in the operation of multiple semiconductor laser diodes. Experimental results show that the side channels can enormously degrade security performance of QKD systems. An important system issue for the improvement of quantum bit error rate (QBER) related with laser driving condition is furtherly addressed with experimental results.
\end{abstract}

\ocis{(270.5568) Quantum cryptography; (060.5565) Quantum communication; (140.5960) Semiconductor lasers.} 


\section{Introduction}
 
Trials of secure information transfer between two distant parties have driven the development of the field of quantum key distribution (QKD) whose security is guaranteed by the nature of quantum physics. Sender and receiver, normally called Alice and Bob, can notice the existence of an unauthenticated eavesdropper, Eve, by investigating quantum bit error rate (QBER) of some fractions of their distilled keys. However, unconditional security of QKD is only guaranteed under the assumption of perfect implementations. In practical systems, many unnoticed degree of freedoms, so called side channels, usually open loopholes for eavesdroppers, which degrades the performance of secret key exchanges \cite{nauerth2009information, lydersen2010hacking, rau2015spatial, ko2016informatic, nakata2016intensity}.
 
Polarization is one of the most prevalent physical observables utilized in the implementations of BB84 \cite{BB84} QKD protocol especially for free-space links \cite{nauerth2013air, wang2013direct}. In most polarization-based QKD systems, Alice randomly turns on one of four semiconductor laser diodes for each time slot to generate one of four polarization quantum states. Previously, several loopholes possibly occurred in the configuration of laser diodes have been studied. In \cite{nauerth2009information}, physical quantities such as spatial, temporal, and spectral characteristics which are possibly different among multiple laser diodes were discussed, which need to be identical to avoid information leakage. Also, side information caused by the disparity of spatial mode due to misalignment among multiple lasers was addressed in \cite{rau2015spatial}. Very recently, intensity fluctuation of a single semiconductor laser diode was reported in periodic pulse generation with a single laser for phase-encoding QKD system \cite{nakata2016intensity}.  
 
In this paper, we firstly report how random bit generation with multiple semiconductor laser diodes opens critical side channels in a polarization-based QKD system. In a semiconductor laser diode, the level of driving current and initial carrier density are key parameters which directly define the shape of output pulses. Contrary to periodic pulse generation, aperiodic pulses by random bit generation with multiple laser diodes result in different initial carrier density conditions for each pulse. It causes two side channels, which are temporal disparity and intensity fluctuation among output pulses. We discuss how these phenomena threat the security performance of QKD systems based on simulation and experimental results. Moreover, we furtherly address an important operation issue of multiple laser diodes in terms of QBER performance in QKD systems.
 
While previously discussed loopholes by utilization of multiple laser diodes were focused on discriminating four laser diodes \cite{nauerth2009information, rau2015spatial}, side information covered in this paper is provided by correlations between pulses coming from the same laser, even though Alice turns on one of four lasers with true randomness. This characteristics unavoidably appears in QKD systems utilizing multiple semiconductor lasers with direct modulation.
 
The remaining part of this paper is structured as follows. In sect. 2, configuration of polarization-based free-space QKD system and operational principles of semiconductor laser focusing on carrier density and photon density are described. In sect. 3, mechanisms of output pulse variations by random bit generation with multiple lasers are discussed with simulation and experimental results. Also, diverse aspects of information leakage caused by the aforementioned loophole are described. An operation issue of multiple laser diodes considering system performance is furtherly addressed in sect.4. Finally, we finish this paper in sect 5. with some concluding remarks.
 
\section{Polarization-based free-space BB84 QKD system}
 
\subsection{General configuration}

\begin{figure*}
\centering
\includegraphics[width=120mm]{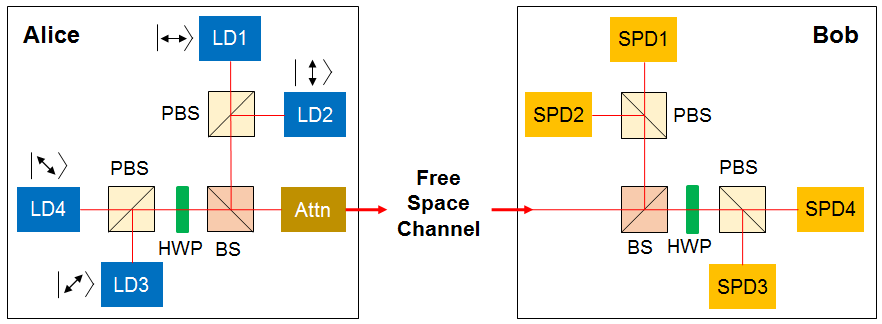}
\caption{\label{fig:BB84setup} Simplified schematic diagram of a free-space BB84 QKD system. LD; laser diode, PBS; polarization beam splitter, BS; beam splitter, HWP; half wave plate, Attn.; attenuator, SPD; single photon detector. }
\end{figure*}

Schematic diagram of a practical free-space BB84 QKD system is shown in Fig.~\ref{fig:BB84setup}. Alice randomly turns on one of four lasers for each time slot to generate polarized photon pulses which are passively created through optical components such as polarizing beam splitter, beam splitter, and half-wave plate. The polarized photon pulses are transmitted to Bob through a free-space channel after they attenuated to the single photon level. Each polarization state is passively decoded as it passes through the Bob's optical system.
 
In a polarization-based BB84 QKD system, four polarization states are generated from four different laser diodes, whereas only single laser diode with a phase modulator is utilized in phase-encoding QKD systems. This is because polarization states used in BB84 protocol are technically much easier to be configured with passive optics, not with a polarization modulator. In decoy BB84 QKD systems, the number of laser diode can be increased to eight or more according to the system configuration \cite{nauerth2009information,peng2007experimental}. 

\subsection{Operation principle of multiple semiconductor laser diodes}

\begin{figure*}
\centering
\includegraphics[width=80mm]{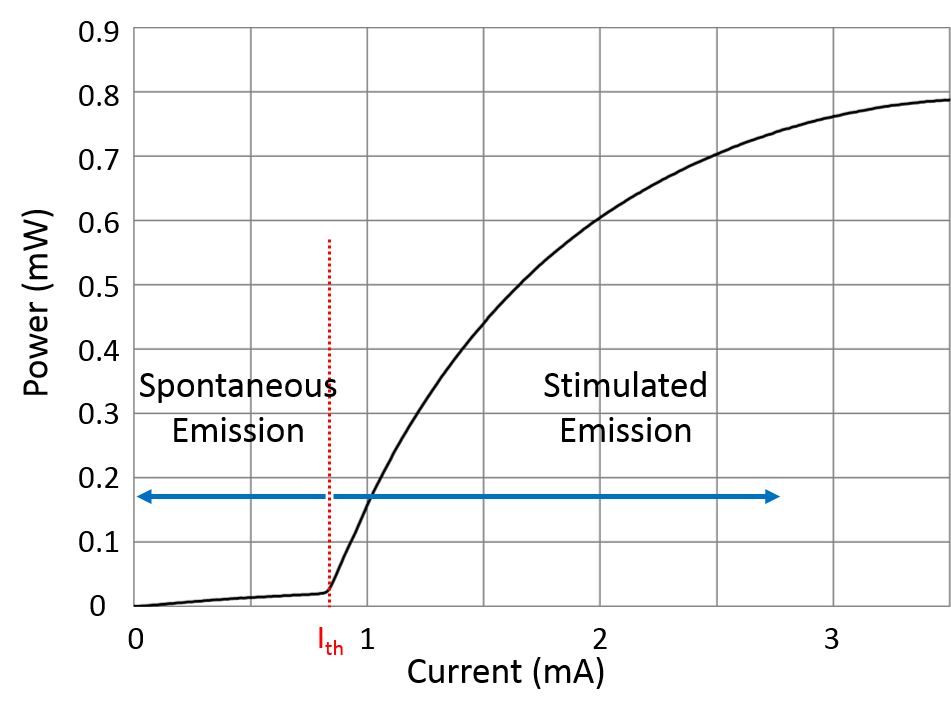}
\caption{\label{fig:LI_curve} Example of L-I curve of a semiconductor laser. }
\end{figure*}

An example of L-I curve of a semiconductor laser is shown in Fig.~\ref{fig:LI_curve}. When driving current is lower than the threshold level $I_{th}$, spontaneous emission is a major process which generates incoherent photons with small output power. If driving current becomes higher than $I_{th}$, stimulated emission starts which generates high power coherent photons. When a laser is under stimulated emission, carrier density goes higher than its threshold level, $N_{th}$, and injected carriers are changed into output photons with certain efficiency. Note that increasing rate of output power under stimulated emission decreases as current increases due to the thermal roll-over effect \cite{colren2012diode} as shown in Fig.~\ref{fig:LI_curve}. 

Short optical pulses are produced by injection of short current pulses into a semiconductor laser. After the injection of current pulses stops, stimulated photon power rapidly decreases and carrier density becomes lower than $N_{th}$. Carrier density keeps decreasing according to the carrier lifetime if no additional driving current pulse is injected. After a period of time much longer than the carrier lifetime, carrier density becomes a steady level, $N_{DC}$, which is determined by the level of DC biased current.

\begin{figure*}
\centering
\includegraphics[width=130mm]{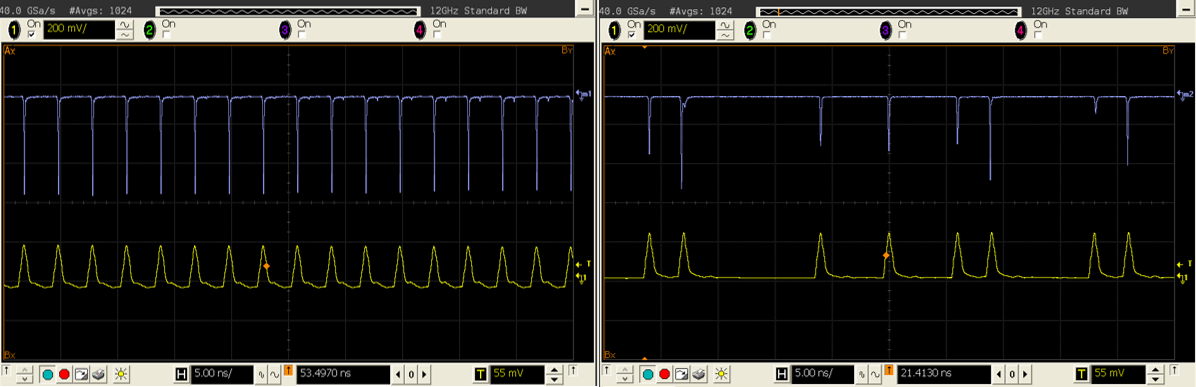}
\caption{\label{fig:Phenomenon} Photon outputs by periodic (left) and aperiodic (right) electrical pulse injection under clock rate of 200MHz with a full width at half maximum (FWHM) of 500ps. Yellow signals in the bottom represents injected electrical pulses and the purple in the upper side represents generated output photon pulses measured with high speed photo-detector with negative polarity. 
}
\end{figure*}

In a phase-encoding BB84 QKD system, a single laser source has nothing to do with random bit generation because quantum phase information is determined by a phase modulator. Thus, we periodically inject same electrical pulses into a single laser to generate identical photon pulses as shown in Fig.~\ref{fig:Phenomenon}(left). On the other hand, in a polarization-based BB84 QKD system using four lasers, each laser diode is switched on with 25\% probability for each time slot to generate random bit information. Thus, each laser diode can be operated in aperiodic manner as shown in Fig.~\ref{fig:Phenomenon}(right). Even though we injected the identical electrical pulses to the same laser, output photon pulses are no longer identical as shown in Fig.~\ref{fig:Phenomenon}(right), which were almost indistinguishable in the periodic case Fig.~\ref{fig:Phenomenon}(left). 

Output pulses are different from one another in two perspectives, which are temporal disparity and intensity fluctuation. Temporal disparity indicates that starting points of output pulses are different with respect to current injection in time domain. Even though it is not clearly seen in the Fig.~\ref{fig:Phenomenon}, one can see the phenomenon in simulated and experimental results in the following sections. Another point, intensity fluctuation among output pulses, is easily seen in the Fig.~\ref{fig:Phenomenon}. These two behaviors among output pulses open unnoticed side channels in QKD systems, which must be properly compensated to guarantee the security.

\section{Behavior of photon pulses caused by random bit generation}


\subsection{Behavior of output photon pulses}

\begin{figure*}
\centering
\includegraphics[width=120mm]{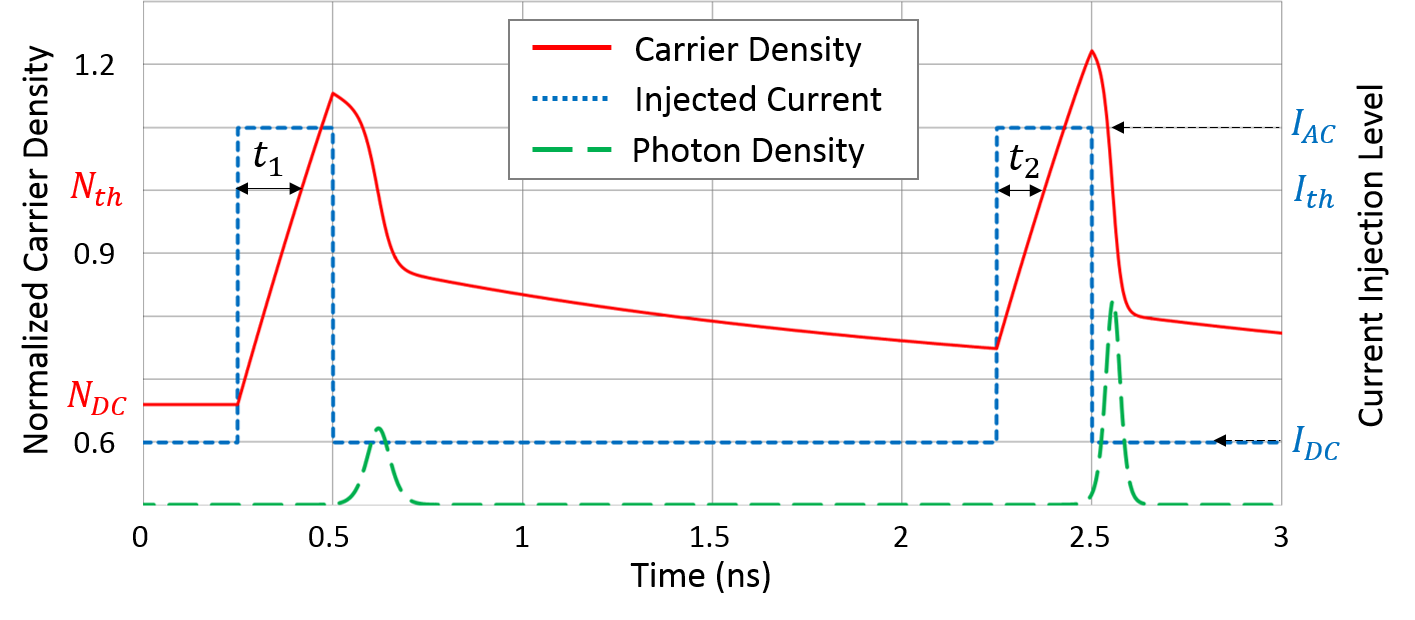}
\caption{\label{fig:Simulation} Simulation result of carrier density (red solid line) and photon density (green dashed line) according to pulse current $I_{AC}$ with DC bias current $I_{DC}$ (blue dotted line). $N_{th}$ is threshold carrier density and $N_{DC}$ is carrier density of DC bias current $I_{DC}$. Common parameters for vertical-cavity surface-emitting laser (VCSEL) are used \cite{colren2012diode}.}
\end{figure*}

Even though Alice drives the same level of current pulses to the same laser diode, temporal behavior of the output pulses can be varied if the initial carrier density is different. Dynamics of carrier and photon density can be simulated with the following rate equations of semiconductor laser diode using common parameters \cite{colren2012diode}. 

\begin{equation}
\frac{dN}{dt} = \frac{\eta_{i}I}{qV} - (R_{sp}+R_{nr}) - v_{g}gN_{p},								\label{eq:carrier_density}
\end{equation}
\begin{equation}
\frac{dN_p}{dt} = \Big(\Gamma v_{g}g - \frac{1}{\tau_{p}}\Big)N_p + \Gamma R_{sp}^{'},			\label{eq:photon_density}
\end{equation}
where $N$ is carrier density, $N_p$ is photon density, $I$ is current, $q$ is electrical charge, and $\tau_p$ is photon lifetime. $R_{sp}, R_{nr}$, and $R_{sp}^{'}$ are recombination terms. $\eta_i, V, \Gamma, g$, and $v_g$ are material parameters of a laser diode. 

Simulation result of the rate equations is shown in Fig.~\ref{fig:Simulation} with the assumption of the steady initial carrier density, $N_{DC}$. Here, Alice generates current pulses of $I_{AC}$ with bias current $I_{DC}$ as described in the Fig.~\ref{fig:Simulation}. As the first current pulse is injected, carrier density, the solid line, reaches to $N_{th}$ after $t_1$ and stimulated emission process begins, which results a photon pulse after a short time. After the AC current pulse disappears, carrier density rapidly decreases, which stops stimulated emission. In this region, carrier density decreases down to the steady level of DC point, $N_{DC}$, after a period of time much longer than the carrier lifetime. If the next electrical pulse injection occurs after the carrier density becomes the steady level, output pulse will be identical with the previous one.  

However, the situation changes if the following electrical pulse is injected before the carrier density reaches down to the steady level as shown in Fig.~\ref{fig:Simulation}. We injected the second current pulse after 2ns from the first one, which drives carrier density to rise again. Here, initial carrier density is somewhat higher than the steady level, $N_{DC}$, as shown in the Fig.~\ref{fig:Simulation}. In this case, it takes a shorter time for carrier density to reach $N_{th}$ $(t_2 < t_1)$, which makes stimulated emission process occur earlier. It causes output pulse to appear earlier than the previous one with respect to current injection time. Moreover, the power of the output pulse is relatively stronger than the previous one because carrier density rises relatively higher. 

Aforementioned phenomena may not be a problem if lasers are operated in a periodic manner because initial carrier density can be same for each repetition even if it is somewhat higher than the $N_{DC}$ level. Also, even in aperiodic operations, temporal disparity and intensity fluctuation could not appear under low-speed systems where minimum time interval between two pulses is much longer than the carrier lifetime. However, under aperiodic operation in high-speed systems, as shown in the Fig.~\ref{fig:Phenomenon}, initial carrier densities are different according to the time interval between two consecutive pulses, which causes temporal disparity and intensity fluctuation. If the time interval is short, initial carrier density for the following pulse is relatively high whereas it would be relatively low if the time interval becomes longer. Due to this mechanism, random bit generation by multiple laser diodes can cause temporal disparity and intensity fluctuation in high-speed QKD systems, which opens side channels for potential eavesdroppers. 




\subsection{Experimental setup}

\begin{figure*}
\centering
\includegraphics[width=120mm]{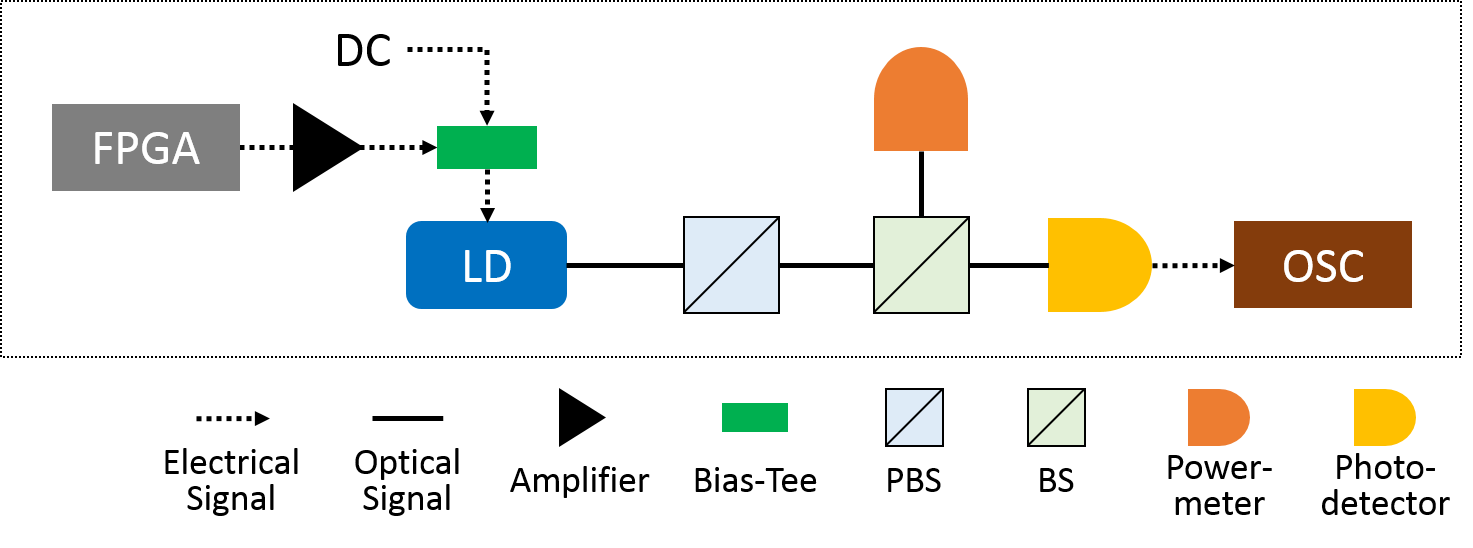}
\caption{\label{fig:setup1} Experimental setup for measuring side channel effects. AC current pulse and DC bias current are injected to a VCSEL laser diode to generate photon pulses which measured at photo-detector and oscilloscope (OSC). Polarization beam splitter (PBS) and beam splitter (BS) are added to make the system similar to a transmitter unit in QKD system. LD; laser diode.}
\end{figure*}

Experimental setup for measuring the side channel effects is shown in Fig.~\ref{fig:setup1}. A polarization beam splitter (PBS) and a beam splitter (BS) are included in order to make the setup similar with the transmitter of polarization-based BB84 QKD system. We adopted a single longitudinal mode VCSEL with lasing wavelength of 787nm. We generated electrical AC current pulses of 500ps duration with the level of $I_{AC} = 4I_{th}$ (FWHM) and three different levels of DC bias current $I_{DC} \in \{0,0.6I_{th},0.9I_{th}\}$. AC and DC signals are combined with a wideband bias-tee. AC pulses are created by a FPGA system and properly amplified to meet the level of pulse current $I_{AC}$. The temporal shape of output pulses is measured by a photo-detector of 12GHz bandwidth and a oscilloscope (OSC) of 13GHz bandwidth. Also, half of the output power split at BS is monitored with a free-space type power-meter with the resolution of 100pW power.

To investigate how time interval between two consecutive pulses impacts on the second output pulses, initial carrier density of the first pulse must be equally maintained because it affects the dynamics of the second pulses. Thus, we set repetition time block as 1us to make the initial carrier density to be in a steady level $N_{DC}$. Note that 1us is much longer than the carrier lifetime of InGaAs/GaAs lasers \cite{colren2012diode} used in the experiments. Under this condition, we can safely ensure that carrier density becomes $N_{DC}$ for each iteration. Within the time block of 1us, we generated two consecutive current pulses with different time intervals from 2ns to 40ns and monitored the output behavior of the second pulses.


\subsection{Experimental results}

\begin{figure*}
\centering
\includegraphics[width=117mm]{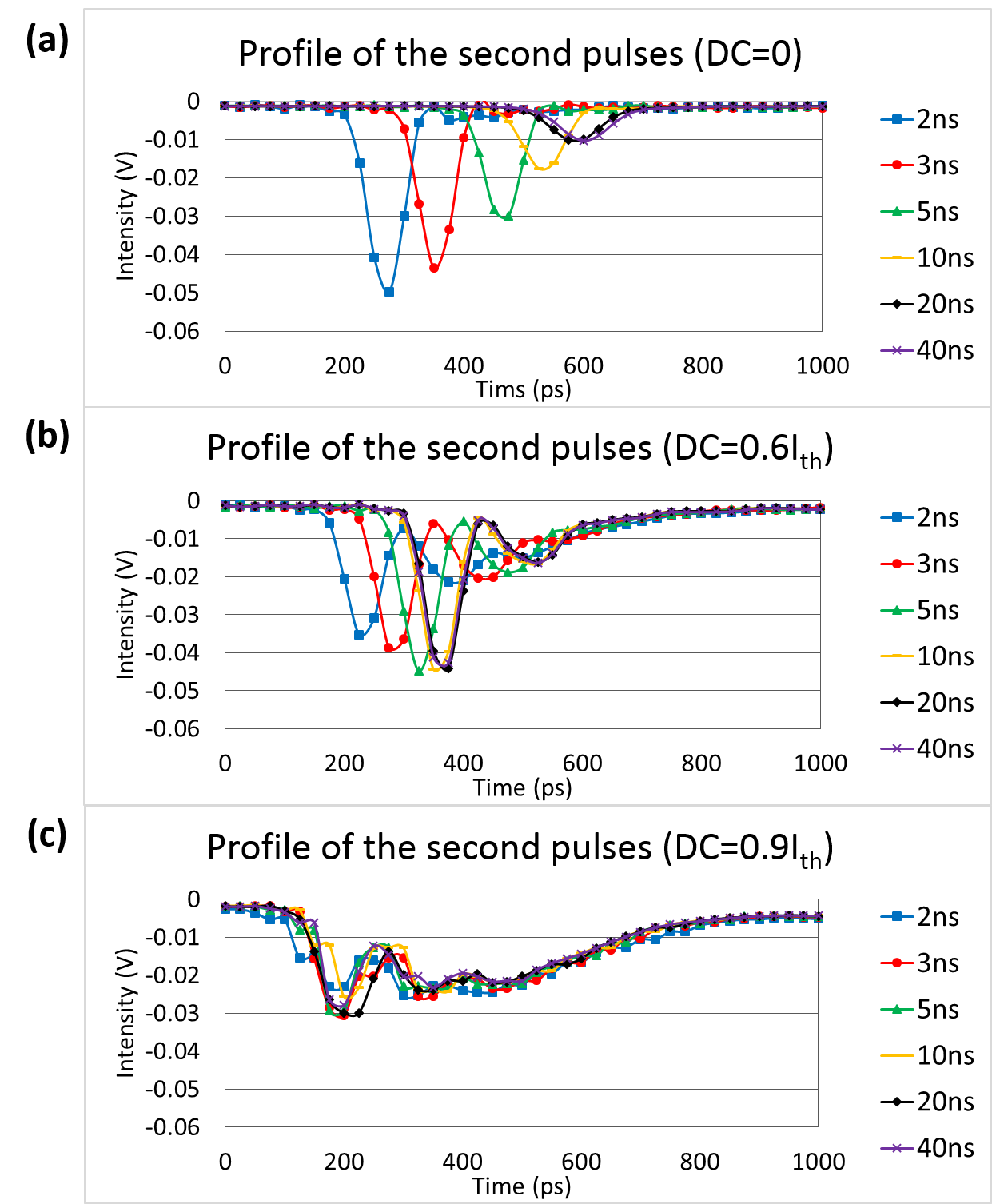}
\caption{\label{fig:results1} Profiles of the second pulses for different time intervals. Output pulses are redrawn with respect to current pulse injection to clearly investigate time delay between current injection and output pulses. (a), (b), and (c) are experimental results for DC = 0, DC = 0.6$I_{th}$, and DC = 0.9$I_{th}$, respectively. Photo-detector generates output pulses with negative polarity. Statistical average of 1024 waveforms is depicted.}
\end{figure*}

Behaviors of output pulses in time domain are shown in Fig.~\ref{fig:results1}. Profiles of the second pulses for DC=0, DC=0.6$I_{th}$, and DC=0.9$I_{th}$ are shown in Fig.~\ref{fig:results1}(a), ~\ref{fig:results1}(b), and ~\ref{fig:results1}(c), respectively. Within the order of carrier lifetime, second pulses are generated in different time position with respect to the current injection time for DC=0 and DC=0.6$I_{th}$. Especially for DC=0 case, output pulses are temporally distinguishable when the time interval is smaller than the 20ns, which corresponds to 50MHz or higher speed operations. Moreover, pulse intensity becomes evidently smaller as time interval becomes longer. Temporal disparity and intensity fluctuation seems to be negligible after 20ns interval, which corresponds to 50MHz or lower speed systems.

Temporal disparity and intensity fluctuation are diminished as DC bias level increases as shown in Fig.~\ref{fig:results1}(b) and Fig.~\ref{fig:results1}(c). This is because that $N_{DC}$ level increases as DC level rises, which curtails time for carrier density to reach down to $N_{DC}$. In case of DC=0.6$I_{th}$, the phenomena become very weak after 10ns interval, which corresponds to 100MHz or lower speed operation. For DC=0.9$I_{th}$, temporal disparity and intensity fluctuation among the second pulses are hardly observed even in the short time interval of 2ns. Note that all photon pulses regardless of time intervals must be temporally overlapped to close the temporal side channels. 

\begin{figure*}
\centering
\includegraphics[width=90mm]{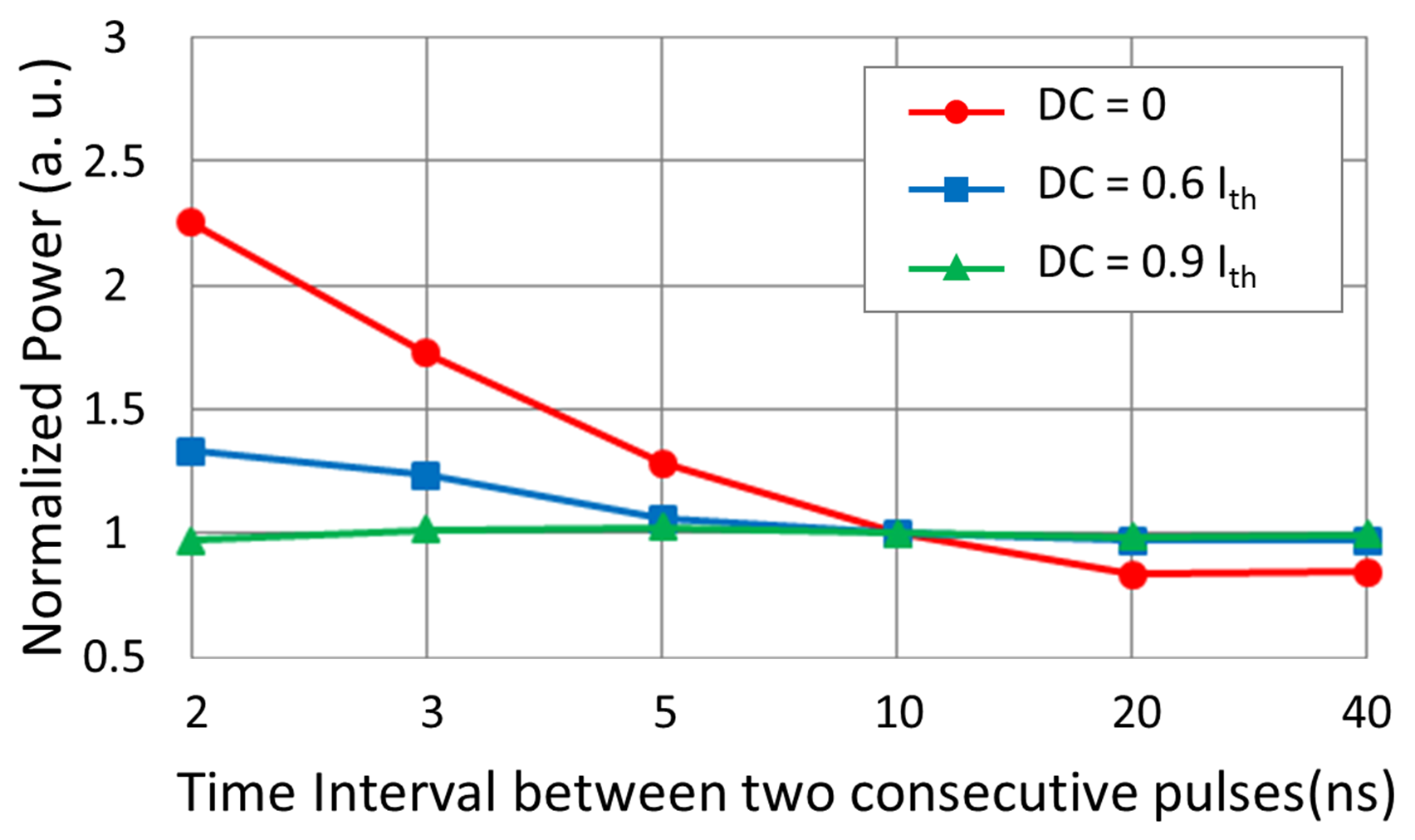}
\caption{\label{fig:results2} Normalized power of the second pulses for different time intervals at different levels of DC bias current.}
\end{figure*}

Intensity of the second pulses was also measured  by a power-meter as shown in Fig.~\ref{fig:results2}. We normalized output pulse powers with the one at 10 ns time interval. Optical power caused by spontaneous emission and the first pulse output are subtracted to properly measure the second output pulses exclusively. For a given DC bias current, intensity of the second pulse decreases as time interval increases. For a given time interval, intensity fluctuation evidently decreases as DC bias current goes close to threshold current level. In case of DC = 0, intensity fluctuation becomes within the range of $\pm$2.5\% for the time intervals longer than 20ns. For DC=0.6$I_{th}$, it becomes within the same range for time intervals longer than 10ns. Intensity fluctuation was almost negligible under DC=0.9$I_{th}$. 


\subsection{Security threat}

Aforementioned phenomena can be extremely critical in terms of security of QKD. The most critical side information is correlation between consecutive pulses possibly revealed by investigating time position of each photon. For example, in case of Fig.~\ref{fig:results1}(a), if a photon is detected in the time position around 250ps, Eve can know that the current pulse was injected to the same laser diode 2ns earlier with high probability. Thus, Eve can guess that the same polarization state must be generated at 2ns earlier. Such correlation between consecutive pulses caused by this behavior destroys randomness even if Alice generates quantum states with true randomness. Under this condition, unambiguous state discrimination (USD) \cite{Dušek2000Unambiguous} attack becomes possible even for the single photon states because Eve can collect the same polarization states from other time positions. Note that USD attack is only valid for the multi-photon states. Also, photon number splitting (PNS) \cite{brassard2000limitations} attack becomes even powerful because Eve can attain additional correlated bits in other time position by investigating a single multi-photon state. Thus, single photon states guaranteed to be secure by decoy methods \cite{hwang2003quantum,lo2005decoy} can be no longer safe, which enormously degrades security performance of QKD systems.

Intensity fluctuation is also a potential threat as already discussed in \cite{nakata2016intensity,wang2008general,hayashi2014security} especially for decoy-BB84 QKD protocol. In \cite{wang2008general}, general theory of security performance of QKD under intensity fluctuation is described with assumption that Eve can knows the relatively strong and weak pulses. It claims that secure key rate described in \cite{peng2007experimental} can be decreased to 70.8 bits/s under intensity error upper bound of 5\%, which was originally 136.3 bits/s. Security analysis with intensity fluctuation has been studied \cite{hayashi2014security,mizutani2015finite} in diverse situations because it can seriously decrease secure key rates. 


\section{Laser operation considering QKD performance}

\begin{figure*}
\centering
\includegraphics[width=110mm]{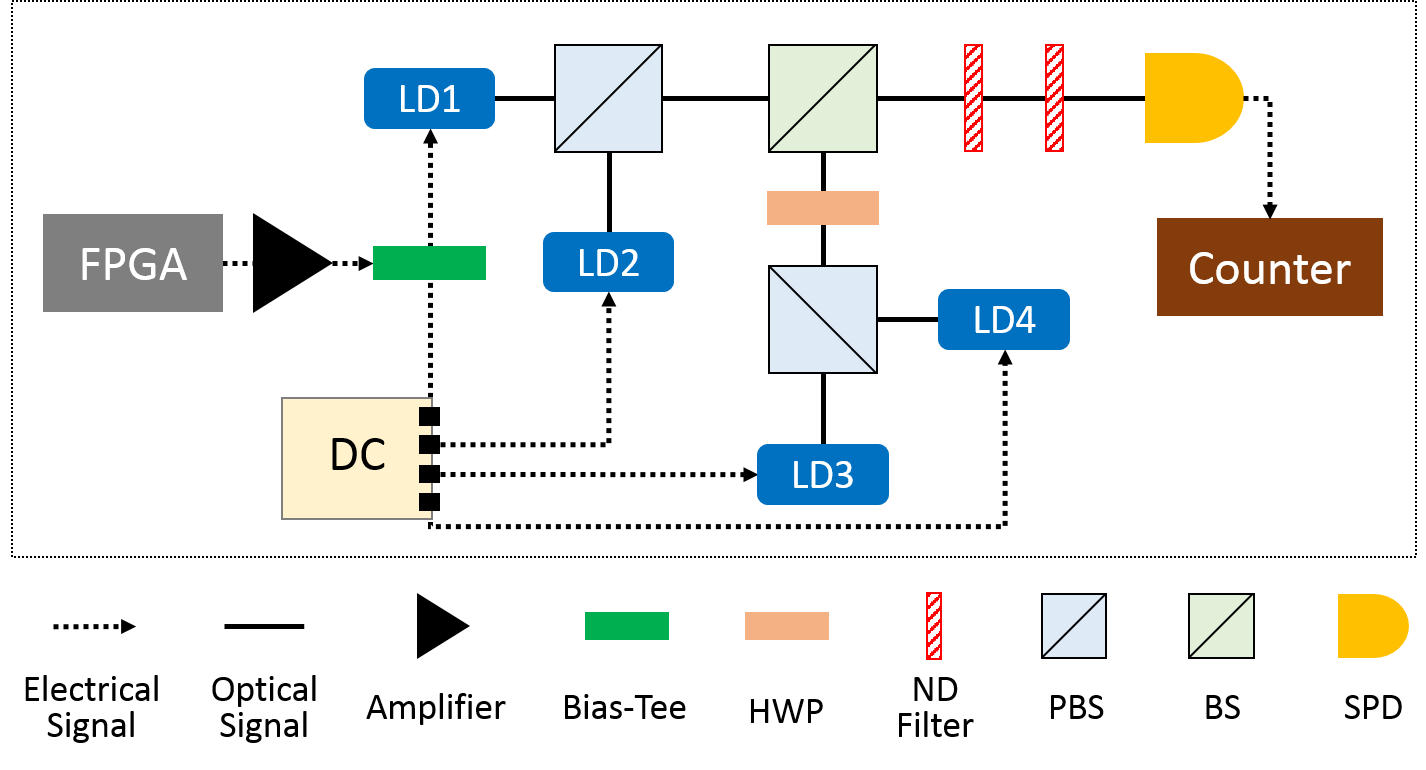}
\caption{\label{fig:setup2} Proof of concept experiment for estimating effects of DC current biased laser diode on the performance of QKD system. LD; laser diode, PBS; polarization beam splitter, BS; beam splitter, HWP; half wave plate, ND; neutral density, SPD; single photon detector.}
\end{figure*}

To guarantee the unconditional security, all photon pulses must be temporally overlapped with same intensity. It seems that condition of high DC bias current improves the quality of temporal overlap and intensity fluctuation as shown in Fig.~\ref{fig:results1} and Fig.~\ref{fig:results2}. However, background photons caused by spontaneous emission process become non-negligible as DC bias current increases as shown in Fig.~\ref{fig:LI_curve}, which can directly increase QBER of QKD systems. 

We estimated this negative effect of spontaneous emission on the performance of BB84 QKD system as shown in Fig.~\ref{fig:setup2}. Four polarization sources of a BB84 QKD system were implemented using four VCSELs with the single longitudinal mode at the wavelength of 787nm. Laser diode1 is operated under the condition of bias current DC=0.9$I_{th}$, pulse current AC=4$I_{th}$ with 500ps (FWHM), and clock rate of 100 MHz with 0.6 mean photon number per pulse. Note that decoy-BB84 QKD system normally uses mean photon number per pulse around 0.6 photon for signal states \cite{wang2013direct,peng2007experimental}. Other three lasers are biased at DC=0.9$I_{th}$ without injection of AC current pulses. Additional ND filter of 18.5dB is adopted for channel loss. Detection outputs are counted using a Si-APD based single photon detector (PerkinElmer SPCM-AQ4C). We repeated the experiment without AC current pulsesinjection under the condition of bias current DC=0.9$I_{th}$, DC=0.6$I_{th}$, and DC=0 exclusively to estimate photon counts caused by the DC bias. The experiments were performed in a dark room condition to eliminate noise by other photons. 

\begin{figure*}
\centering
\includegraphics[width=100mm]{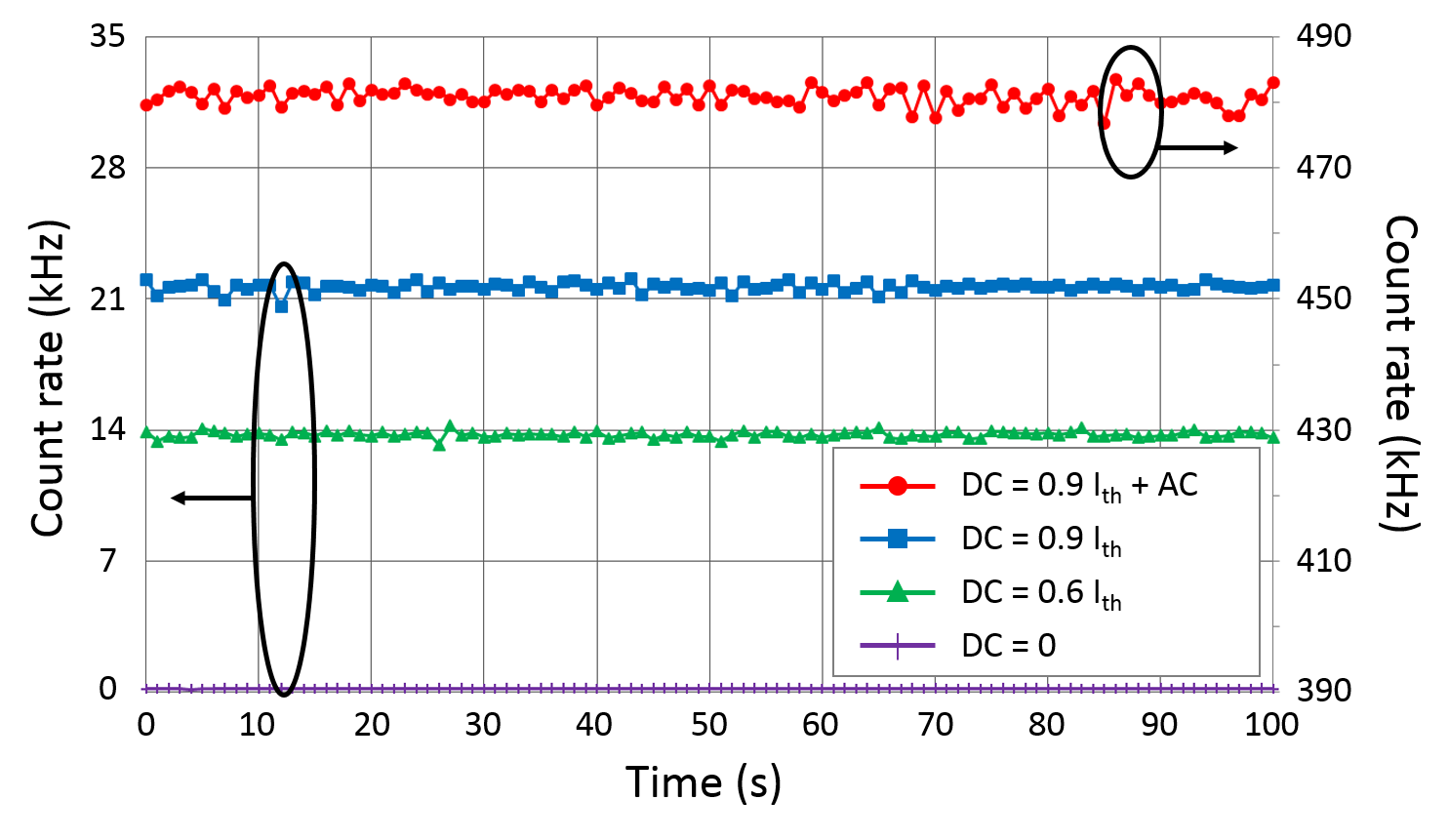}
\caption{\label{fig:results3} Single photon counting results of QKD system and DC bias level. }
\end{figure*}

Experimental results are measured for 100 seconds as shown in Fig.~\ref{fig:results3}. Count rates of 481kHz (Avg.: 480,924 counts/s, Std. Dev.: 1347) and 22kHz (Avg.: 21,642 counts/s, Std. Dev.: 242) are recorded under the condition of DC=0.9$I_{th}$+AC and DC=0.9$I_{th}$, respectively. Thus, the count rate of 22kHz out of 481kHz can be roughly interpreted as photon counts caused by spontaneous emission under DC=0.9$I_{th}$, which spread in all time domain. We measured photon counts for the case of DC=0.6$I_{th}$ and DC=0 for comparison which are 14kHz (Avg.: 13,756 counts/s, Std. Dev.: 161) and 180Hz (Std. Dev.: 12), respectively. Count rate at DC=0 indicates dark count rate of the single photon detector.  
 
Noise photon count of 22kHz caused by spontaneous emission is not negligible compared to the signal count of 481kHz. It can definitely increase the QBER of the system, which directly decreases secret key rates. If a system utilizes eight or more laser diodes to configure the decoy states \cite{nauerth2009information,peng2007experimental}, the noise count can become double or more. Considering the fact that polarization-based systems are mostly used for free-space channel where other photon noises exist including the sunlight, noise effect caused by the source itself must be minimized to improve the system performance. Thus, high DC bias level may not be a good solution for the performance of QKD systems. 

Steps for finding desirable DC bias condition considering both information leakage and system performance are as follows. First, one needs to figure out minimum time interval between two consecutive pulses from a single laser diode, which is determined by the system clock rate. Second, one should investigate minimum DC bias current level eliminating temporal disparity and intensity fluctuation among the pulses from the minimum time interval to longer ones. Third, one must check whether QBER caused by noise photon counts due to DC bias current is negligible or not. If negative effects caused by the DC bias current is non-negligible, system operation should be slow down to decreases the DC bias level condition. Optimal DC bias current for a fixed system clock rate can be different according to the structure of the semiconductor laser diode and its material characteristics. 




\section{Conclusion}

In this paper, we discussed critical side channel effects possibly occurred in random bit generation with multiple semiconductor laser diodes, which are temporal disparity and intensity fluctuation among photon pulses. We showed that the phenomena are severe under low DC bias condition, which allows an eavesdropper to obtain correlations between consecutive pulses from the same laser diode. The negative effects become worse for the high-speed operation where time interval between two consecutive pulses is within the order of carrier lifetime of the laser diode. The information leakage caused by the behaviors could be alleviated as DC bias current increases close to the threshold level. However, we furtherly addressed that QBER performance can be degraded under high DC current situation due to background photons caused by spontaneous emission, which almost linearly increases with DC current level. Thus, AC pulse and DC bias current must be elaborately controlled with the system operation speed, considering both information leakage and system performance. 


\section*{Acknowledgments}
This work was supported by Electronics and Telecommunications Research Institute (ETRI) grant funded by the Korean government. [Development of preliminary technologies for transceiver key components and system control in polarization based free space quantum key distribution]



\begin{thebibliography}{99}

\bibitem{lydersen2010hacking}
Lydersen, L., Wiechers, C., Wittmann, C., Elser, D., Skaar, J., and Makarov, V., "Hacking commercial quantum cryptography systems by tailored bright illumination", Nat. Photon. \textbf{4}, 686-689 (2010).

\bibitem{ko2016informatic}
Ko, H., Lim, K., Oh, J., and Rhee, J. K. K., "Informatic analysis for hidden pulse attack exploiting spectral characteristics of optics in plug-and-play quantum key distribution system", Quant. Inf. Proc. \textbf{15}, 4265-4282 (2016).

\bibitem{nauerth2009information}
Nauerth, S., Furst, M., Schmitt-Manderbach, T., Weier, H., and Weinfurter, H., "Information leakage via side channels in freespace BB84 quantum cryptography", New J. Phys. \textbf{11}, 065001 (2009).

\bibitem{rau2015spatial}
Rau, M., Vogl, T., Corrielli, G., Vest, G., Fuchs, L., Nauerth, S., and Weinfurter, H., "Spatial mode side channels in free-space QKD implementations", IEEE J. Sel. Top. Quantum Electron. \textbf{21}, 187-191 (2015).

\bibitem{nakata2016intensity}
Nakata, K., Tomita, A., Fujiwara, M., Yoshino, K. I., Tajima, A., Okamoto, A., and Ogawa, K., "Intensity fluctuation of a gain-switched semiconductor laser for quantum key distribution systems", Opt. Express \textbf{25}, 622-634 (2017).

\bibitem{BB84}
Bennett. C. H, and Brassard. G., "Quantum cryptography: Public key distribution and coin tossing", in Proceedings of IEEE International Conference on Computers, Systems, and Signal Processing, pp. 175-179 (1984).

\bibitem{nauerth2013air}
Nauerth, S., Moll, F., Rau, M., Fuchs, C., Horwath, J., Frick, S., and Weinfurter, H., "Air-to-ground quantum communication", Nat. Photon. \textbf{7}, 382-386 (2013).

\bibitem{wang2013direct}
Wang, J. Y., Yang, B., Liao, S. K., Zhang, L., Shen, Q., Hu, X. F.,  Wu, J. C., Yang, S. J., Jiang, H., Tang, Y. L., Zhong, B., Liang, H. Liu, W. Y., Hu, Y. H., Huang, Y. M., Qi, B., Ren, J. G., Pan, G. S., Yin, J. Jia, J. J., Chen, Y. A., Chen, K., Peng, C. Z., and Pan, J. W., "Direct and full-scale experimental verifications towards ground-satellite quantum key distribution", Nat. Photon. \textbf{7}, 387-393 (2013).

\bibitem{peng2007experimental}
Peng, C. Z., Zhang, J., Yang, D., Gao, W. B., Ma, H. X., Yin, H., Zeng, H. P., Yang, T., Wang, X. B., and Pan, J. W., "Experimental long-distance decoy-state quantum key distribution based on polarization encoding", Phys. Rev. Lett. \textbf{98}, 010505 (2007).

\bibitem{colren2012diode}
Coldren, L. A., Corzine, S. W., and Mashanovitch, M. L., Diode lasers and photonic integrated circuits (John Wiley and Sons, 2012).

\bibitem{Dušek2000Unambiguous}
Dusek, M., Jahma, M., and Lutkenhaus, N., "Unambiguous state discrimination in quantum cryptography with weak coherent states", Phys. Rev. A \textbf{62}, 022306 (2000).

\bibitem{brassard2000limitations}
Brassard, G., Lutkenhaus, N., Mor, T., and Sanders, B. C., "Limitations on practical quantum cryptography", Phys. Rev. Lett. \textbf{85}, 1330 (2000).

\bibitem{hwang2003quantum}
Hwang, W. Y., "Quantum key distribution with high loss: toward global secure communication", Phys. Rev. Lett. \textbf{91}, 057901 (2003).
 
\bibitem{lo2005decoy}
Lo, H. K., Ma, X., and Chen, K., "Decoy state quantum key distribution", Phys. Rev. Lett. \textbf{94}, 230504.(2005).

\bibitem{wang2008general}
Wang, X. B., Peng, C. Z., Zhang, J., Yang, L., and Pan, J. W., "General theory of decoy-state quantum cryptography with source errors", Phys. Rev. A \textbf{77}, 042311 (2008).

\bibitem{hayashi2014security}
Hayashi, M., and Nakayama, R., "Security analysis of the decoy method with the Bennett-Brassard 1984 protocol for finite key lengths", New J. Phys. \textbf{16}, 063009 (2014).

\bibitem{mizutani2015finite}
Mizutani, A., Curty, M., Lim, C. C. W., Imoto, N., and Tamaki, K., "Finite-key security analysis of quantum key distribution with imperfect light sources", New J. Phys. \textbf{17}, 093011 (2015). 

\end{thebibliography}
\end{document}